\documentclass[twocolumn,aps,prb]{revtex4-1}

\usepackage{amsmath}
\usepackage{graphicx}
\usepackage{color}
\usepackage{url}
\usepackage[left=0.75in, right=0.75in,top=0.75in,bottom=1.5in]{geometry}

\begin{document}

\title{Micromagnetic simulations with periodic boundary conditions: Hard-soft nanocomposites}

\author{Aleksander L. Wysocki}
\email{alexwysocki2@gmail.com}

\author{Vladimir P. Antropov}

\affiliation{Ames Laboratory, Ames, IA 50011, USA}

\date{\today}

\begin{abstract}
We developed a micromagnetic method for modeling magnetic systems with periodic boundary conditions along an arbitrary number of dimensions. The main feature is an adaptation of the Ewald summation technique for evaluation of long-range dipolar interactions. The method was applied to investigate the hysteresis process in hard-soft magnetic nanocomposites with various geometries. The dependence of the results on different micromagnetic parameters was studied. We found that for layered structures with an out-of-plane hard phase easy axis the hysteretic properties are very sensitive to the strength of the interlayer exchange coupling, as long as the spontaneous magnetization for the hard phase is significantly smaller than for the soft phase. The origin of this behavior was discussed. Additionally, we investigated the soft phase size optimizing the energy product of hard-soft nanocomposites.
\end{abstract}

\maketitle

\section{Introduction}

One of the biggest challenges in micromagnetic modeling\cite{Berkov,Fidler,Kronmuller} is a proper treatment of the dipolar coupling. Due to the long-range character of this interaction, an accurate evaluation of the corresponding energy is a notoriously expensive computational task even for finite systems. The problem becomes more severe when periodic boundary conditions (PBC) are used. In this case, a supercell is introduced which is repeated infinitely many times along one, two, or three directions forming an infinite system. A magnetic moment inside the supercell interacts not only with other magnetic moments inside the supercell but also with magnetic moments inside all periodic images of the supercell. 

Early solution to the problem involved neglecting the interaction beyond some cutoff radius.\cite{Zhu} Similar idea is employed in the more recently proposed macrogeometry method.\cite{Fangohr} Another approach is based on the Lorenz cavity concept.\cite{Vos} Many authors evaluated the dipolar interactions using the convolution theorem and fast Fourier transform (FFT) method.\cite{Kuma} For an infinite range of interactions, however, the convolution theorem is not valid and errors may be expected when magnetization varies rapidly in space.\cite{Berkov2} For one-dimensional (1D) and two-dimensional (2D) PBC it was shown that the dipolar interactions can be accurately evaluated by using analytical formulas for large distances.\cite{Lebecki,Wang} This method, however, has not been extended to the three-dimensional (3D) case. The most natural approach for calculations of the long-range interaction with PBC is based on the Ewald summation.\cite{Ewald,Leeuw} However, this technique is rarely used in micromagnetic simulations. A notable exception is the work in Ref. \onlinecite{Berkov2} where the Ewald method was described for the 2D system. 

Hard-soft magnetic nanocomposites, also called exchange spring magnets, are a promising class of materials with potential applications in permanent magnetism\cite{Kneller,Skomski} and magnetic recording.\cite{Richter,Victora} Addition of a high-magnetization soft phase (SP) to a high-anisotropy hard phase (HP) is expected to increase the remanence of the system at the expense of the coercivity. However, if the size of the SP is sufficiently small, the coercivity decrease is only mild resulting in an enhancement of the energy product of the system.\cite{Kneller,Skomski} Thus, hard-soft nanocomposites are candidates for a new generation of permanent magnets.\cite{SkomskiBook,Coey,Poudyal,Jiang} On the other hand, reduction of the coercivity in exchange spring magnets is a desirable feature for magnetic recording applications as it allows for lower writing fields without compromising the thermal stability.\cite{Victora2,Suess}

In this paper we introduce a micromagnetic approach for magnetic hysteresis modeling in systems with 1D, 2D, and 3D PBC. The dipolar interactions are efficiently evaluated using the Ewald summation technique. The method was applied to study magnetic hysteresis of hard-soft nanocomposites with different geometries. We pay a special attention to layered structures with an out-of-plane HP easy axis. For such systems it was demonstrated that the SP magnetization tilt away from the hard phase magnetization direction in order to avoid formation of magnetic charges at the interface.\cite{Belemuk,Belemuk4} Here, we investigate this effect in details by varying the HP spontaneous magnetization. This allows us to control the strength of the interfacial magnetic charges. We demonstrate that the SP tilting competes with the interface exchange coupling (IEC) which favors the perfect alignment of the HP and SP magnetizations. As a results, the hysteretic properties show a strong dependence on the IEC, provided the spontaneous magnetization for the HP is significantly smaller than for the SP. Furthermore, we investigate the optimal SP size for which the maximum energy product is the largest.

Micromagnetic methodology is described in Section II. Section III provides a basic background on hard-soft magnetic nanocomposites. The simulation results for layered and core-shell geometries are presented and discussed in Section IV.  The work is summarized in Section V. Derivation of the Ewald summation technique is given in the Appendix.

\section{Methodology}

\begin{figure}[t!]
\includegraphics[width=1.0\hsize]{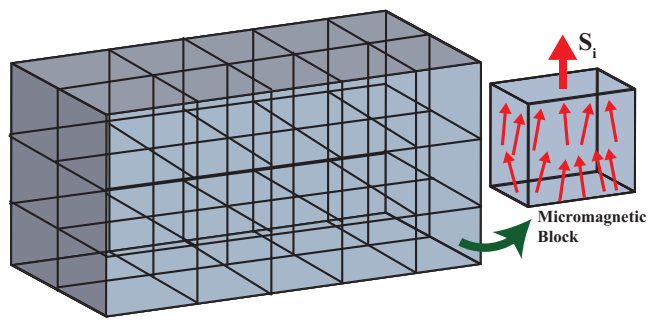}
\caption{Schematic representation of the micromagnetic coarse-grained procedure. A magnetic system is divided into cubic micromagnetic blocks. Atomic magnetic moments (thin red arrows) inside the block $\mathbf{i}$ are approximately parallel and their direction is represented by a unit vector $\mathbf{S}_\mathbf{i}$ called a block spin (thick red arrow).}
\label{MBlocks}
\end{figure}

Micromagnetic methodology can be derived using a coarse-grained procedure which averages out the atomic structure of the system and represents it by a collection of micromagnetic blocks, see Fig. \ref{MBlocks}. Each micromagnetic block contains many atoms whose magnetic moments are assumed to be parallel to each other. Therefore, the size of the micromagnetic blocks must be smaller than the length scale of characteristic magnetic inhomogeneities in the system. The latter is characterized by $l=\text{min}(\delta,l_{ex})$ where $\delta=\pi\sqrt{A/K}$ is the Bloch domain wall width and $l_{ex}=\sqrt{A/(\mu_0M^2)}$ is the exchange length. Here $A$, $K$, and $M$ are the exchange stiffness constant, the anisotropy density constant, and spontaneous magnetization, respectively. For majority of magnets $l$ is of the order of a few nanometers\cite{SkomskiBook} which sets the limit for the micromagnetic block size. However, it should be kept in mind that even if the block size is smaller than $l$, errors are expected in the  description of topological defects such as Bloch points. Additionally, the method breaks down at temperatures close to the Curie point where short-wavelength thermal fluctuations play a significant role. In this case, the fully atomic description is necessary, although an approach based on the Landau-Lifshitz-Bloch equation is an interesting alternative.\cite{Garanin}

The shape of the micromagnetic blocks is arbitrary and can be chosen according to the problem of interest. In addition, both the size and shape of the blocks can vary across the system. In this work, however, we choose all blocks to be cubes with the side length $a$. Note that while this choice requires the system to have a cuboidal shape, more complicated geometries may still be studied by introducing vacuum micromagnetic blocks. We choose the coordinate system to be aligned with the cuboid axes. The size of the system along the Cartesian direction $\alpha$ (in units of $a$) is then denoted by $L_{\alpha}$ ($\alpha=x,y,z$).

Different types of boundary conditions can be adapted. In the case of open boundary conditions the system is surrounded by a vacuum allowing the study of an actual magnetic sample. However, samples of interest often exceed the size limits that are computationally feasible when using micromagnetic calculations. In such cases, a common approach is to consider only a smaller system representing a characteristic microstructure of the magnetic sample. Here, the open boundary conditions introduce artificial surface effects complicating the interpretation of the results. In order to remove the surface effects one can use PBC. In this case, a supercell is introduced which is repeated infinitely many times along one, two, or three directions forming an infinite system. Our approach treats the open boundary conditions as well as 1D, 2D, and 3D PBC on equal footing. In the case of 3D PBC, the system is assumed to be in a closed circuit environment so that there is no macroscopic demagnetization field. We point out, however, that it does not lead to any loss of generality since the hysteresis loop can be appropriately rescaled to take into account any demagnetization field. 

The direction of atomic magnetic moments inside a micromagnetic block is described by a unit vector $\mathbf{S}_\mathbf{i}$ called a block spin (the subscript $\mathbf{i}$ denotes the position of the block). The set of all block spins $\{\mathbf{S}_\mathbf{i}\}$ fully specifies the magnetic state of the system. The magnetic energy depends on the block spin configuration and it consists of four terms

\begin{equation}
E=E_{H}+E_{K}+E_{X}+E_{D}
\label{Energy}
\end{equation}

Here $E_{H}$, $E_{K}$, $E_{X}$, and $E_{D}$ denote Zeeman, anisotropy, exchange, and dipolar contributions to energy, respectively. The Zeeman energy describes the interaction of magnetic moments with an external magnetic field ($\mathbf{H}$) and can be written as

\begin{equation}
E_{H}=-\mu_0V_a\sum_\mathbf{i}M_\mathbf{i}\mathbf{H}\cdot\mathbf{S}_\mathbf{i}
\label{Zeeman}
\end{equation}

where $\mu_0=4\pi 10^{-7}$ N/A$^2$ is the vacuum permeability, $V_a=a^3$ is the micromagnetic block volume, and $M_{\mathbf{i}}$ is the spontaneous magnetization for the block $\mathbf{i}$.

The anisotropy energy describes the preference of magnetic moments to point along their easy axis. We consider a second-order uniaxial expression given by

\begin{equation}
E_{K}=-V_a\sum_{\mathbf{i}}K_\mathbf{i}(\mathbf{\hat{n}}_\mathbf{i}\cdot\mathbf{S}_\mathbf{i})^2
\label{Anisotropy}
\end{equation}

where $K_\mathbf{i}$ is the anisotropy constant density of the block $\mathbf{i}$ and $\mathbf{\hat{n}}_\mathbf{i}$ is the unit vector along its easy axis. More general forms of the anisotropy energy can be easily implemented.

The exchange energy favors neighboring magnetic moments to be aligned. It is given by the Heisenberg formula

\begin{equation}
E_{X}=-\frac{1}{2}\sum_{<\mathbf{i}\mathbf{j}>}J_{\mathbf{i}\mathbf{j}}\mathbf{S}_\mathbf{i}\cdot\mathbf{S}_\mathbf{j}
\label{Exchange}
\end{equation}

Here the summation is over the nearest neighbor blocks and $J_{\mathbf{i}\mathbf{j}}$ is the exchange coupling. Implementation of PBC for the exchange energy is rather trivial since the interaction is short-ranged. 

The dipolar energy corresponds to the magnetostatic interaction between classical dipoles and is given by 

\begin{equation}
E_{D}=-\frac{1}{2}\frac{\mu_0}{4\pi}V_a\sum_{\mathbf{i}\mathbf{j}}{}^{'}M_\mathbf{i}M_\mathbf{j}\mathbf{S}_\mathbf{i}\cdot\mathbf{D}_{\mathbf{i}\mathbf{j}}\cdot\mathbf{S}_\mathbf{j}
\label{Dipolar}
\end{equation}

where the prime excludes from the summation the self-interaction term $\mathbf{i}=\mathbf{j}$ and $\mathbf{D}_{\mathbf{i}\mathbf{j}}$ is the dipolar tensor with Cartesian elements given by

\begin{equation}
D_{\mathbf{i}\mathbf{j}}^{\alpha\beta}=\left(3r_{\mathbf{i}\mathbf{j}}^{\alpha}r_{\mathbf{i}\mathbf{j}}^{\beta}-r_{\mathbf{i}\mathbf{j}}^{2}\delta_{\alpha\beta} \right)/r_{\mathbf{i}\mathbf{j}}^{5}
\label{DipTensor}
\end{equation}

Here $\alpha=x,y,z$ and $r_{\mathbf{i}\mathbf{j}}^{\alpha}$ is the component of a vector pointing from $\mathbf{i}$ to $\mathbf{j}$ in units of $a$. Due to long-range character of dipolar interactions, this contribution to the energy should be treated with special care in the presence of PBC. Indeed, while in Eq. (\ref{Dipolar}) we can choose the index $\mathbf{i}$ to run over micromagnetic blocks in the supercell, the index $\mathbf{j}$ then runs over the blocks inside the supercell and all its periodic images. It is useful to make the substitution $\mathbf{j}\rightarrow\mathbf{j}+\mathbf{n}$ with the new index $\mathbf{j}$ running over the blocks only inside the supercell. Here, $\mathbf{n}=(n_x,n_y,n_z)$ where $n_\alpha$ is an integer enumerating the periodic images of the supercell if PBC are assumed along the direction $\alpha$ and zero otherwise. Eq. (\ref{Dipolar}) can be then written as

\begin{equation}
E_{D}=-\frac{1}{2}\frac{\mu_0}{4\pi}V_a\sum_{\mathbf{i}\mathbf{j}}M_\mathbf{i}M_\mathbf{j}\mathbf{S}_\mathbf{i}\cdot\mathbf{\bar{D}}_{\mathbf{i}\mathbf{j}}\cdot\mathbf{S}_\mathbf{j}
\label{Dipolar2}
\end{equation}

where the summation is over the blocks inside the supercell and we defined the effective dipolar tensor as

\begin{equation}
\mathbf{\bar{D}}_{\mathbf{i}\mathbf{j}}=\sum_{\mathbf{n}}{}^{'}\mathbf{D}_{\mathbf{i},\mathbf{j+\mathbf{n}}}
\label{DipTensorPBC}
\end{equation}

Here the prime excludes from the sum the $\mathbf{n}=\mathbf{0}$ term when $\mathbf{i}=\mathbf{j}$. When open boundary conditions are used, only $\mathbf{n}=\mathbf{0}$ survives making evaluation of (\ref{DipTensorPBC}) trivial. In the case of PBC we evaluate the effective dipolar tensor using the Ewald summation technique.\cite{Ewald} For 3D PBC the application of this technique to the dipolar systems is well known.\cite{Leeuw} The basic idea is to add for each micromagnetic block a Gaussian magnetic charge distribution corresponding to the dipole moments exactly opposite to the original point-like dipole of the block. These Gaussian dipoles screen the resulting dipolar interaction making it short-ranged so that the corresponding summation (\ref{DipTensorPBC}) can be straightforwardly done. The extra field from the smooth Gaussian magnetic charge density can then be easily evaluated in the Fourier space and subtracted. The generalizations of the Ewald technique to the system with 2D and 1D PBC were described in Refs. \onlinecite{Grzybowski,Porto} for the case of electrostatic interactions. The application of the Ewald method for evaluation of the effective dipolar tensor $\mathbf{\bar{D}}_{\mathbf{i}\mathbf{j}}$ is presented in the Appendix. Note that for a given system, the calculation of $\mathbf{\bar{D}}_{\mathbf{i}\mathbf{j}}$ needs to only be done once, therefore it has a negligible contribution to the total computational time. On the other hand, the evaluation of the dipolar energy from Eq. (\ref{Dipolar2}) or the dipolar contribution to the effective field (last term in Eq. (\ref{Heff2})) can be very time consuming for large systems. The fast Fourier transform (FFT) can be used in such cases to save a considerable amount of time. It should be kept in mind, however, that the zero-padding technique needs to be applied along the directions with no PBC. 

The time evolution of block spins is described by the Landau-Lifshitz-Gilbert equation

\begin{eqnarray}
\nonumber
\frac{d\mathbf{S}_\mathbf{i}}{dt}=-\frac{\gamma}{1+\alpha^2}\mathbf{S}_\mathbf{i}\times\mathbf{H}_\mathbf{i}^{\text{eff}}-\frac{\alpha\gamma}{1+\alpha^2}\mathbf{S}_\mathbf{i}\times(\mathbf{S}_\mathbf{i}\times\mathbf{H}_\mathbf{i}^{\text{eff}}) \\
\label{LLG}
\end{eqnarray}

where $\gamma$ is the electron gyromagnetic ratio, $\alpha$ is the Gilbert damping factor, and $\mathbf{H}_{\mathbf{i}}^{\text{eff}}$ is the effective field defined as

\begin{equation}
\mathbf{H}_\mathbf{i}^{\text{eff}}=-\frac{1}{\mu_0M_\mathbf{i}V_a}\frac{\partial E}{\partial\mathbf{S}_\mathbf{i}}
\label{Heff}
\end{equation}

Using Eqs. (\ref{Energy}-\ref{DipTensorPBC}) we obtain

\begin{widetext}
\begin{equation}
\mathbf{H}_\mathbf{i}^{\text{eff}}=\mathbf{H}+H^K_\mathbf{i}(\mathbf{\hat{n}}_\mathbf{i}\cdot\mathbf{S}_\mathbf{i})\mathbf{\hat{n}}_\mathbf{i}+\frac{1}{\mu_0M_\mathbf{i}V_a}\sum_{<\mathbf{j}>}J_{\mathbf{ij}}\mathbf{S}_{\mathbf{j}}+\frac{1}{4\pi}\sum_\mathbf{j}M_\mathbf{j}\mathbf{\bar{D}}_{\mathbf{i}\mathbf{j}}\cdot\mathbf{S}_\mathbf{j}
\label{Heff2}
\end{equation}
\end{widetext}

where $H^K_\mathbf{i}=\frac{2K_\mathbf{i}}{\mu_0M_\mathbf{i}}$ is the anisotropy field of the micromagnetic block $\mathbf{i}$.

The first term in Eq. (\ref{LLG}) describes the precession of the block spins around their effective field direction while the second term is responsible for relaxation of the block spins towards the local or global energy minima. However, if we are not interested in dynamics, but merely in finding a stable state, we can neglect the precessional motion. In this case, discretization of Eq. (\ref{LLG}) results in a following iterative procedure for finding a stable block spin configuration\cite{Berkov}

\begin{equation}
\mathbf{S}_\mathbf{i}^{\text{new}}=\mathbf{S}_\mathbf{i}^{\text{old}}-\tilde\alpha\mathbf{S}_\mathbf{i}^{\text{old}}\times(\mathbf{S}_\mathbf{i}^{\text{old}}\times\mathbf{H}_\mathbf{i}^{\text{eff}})
\label{Alg}
\end{equation}

Here $\mathbf{H}_\mathbf{i}^{\text{eff}}$ is the effective field calculated using $\{\mathbf{S}_\mathbf{i}^{\text{old}}\}$ and the parameter $\tilde\alpha$ is an effective mixing parameter which can be adaptively adjusted during the iterative procedure in order to improve the convergence. Similarly, as in Ref. \onlinecite{Berkov} the new block spin configuration $\{\mathbf{S}_\mathbf{i}^{\text{new}}\}$ is accepted only when it decreases the system energy. Otherwise, $\tilde\alpha$ is reduced and another configuration is tried according to Eq. (\ref{Alg}). This condition ensures the stability of the algorithm. We assume that the stable state is achieved when $|\mathbf{S}_\mathbf{i}\times\mathbf{H}_\mathbf{i}^{eff}|<10^{-3}$ for all sites.

Micromagnetic simulations require knowledge of spontaneous magnetization $M_\mathbf{i}$, anisotropy density constant $K_\mathbf{i}$, and exchange couplings $J_{\mathbf{ij}}$. These parameters are often obtained from experimental data. In particular, for each constituent, $M_\mathbf{i}$ and $K_\mathbf{i}$ can be estimated from the magnetization measurements for the corresponding bulk material.\cite{SkomskiBook} In the case of the exchange coupling $J_{\mathbf{ij}}$, as long as the blocks $\mathbf{i}$ and $\mathbf{j}$ are made of the same material, it can be related to the exchange stiffness of this material ($A_\mathbf{i}$) using $J_{\mathbf{i}\mathbf{j}}=2aA_\mathbf{i}$. The exchange stiffness can be then obtained from neutron diffraction data using the equation $A=DM/(2g\mu_B)$ where $D$ is the measured spin wave stiffness, $g$ is the Land{\'e} g-factor, and $\mu_B$ is the Bohr magneton. The exchange stiffness can be also estimated from the Curie temperature ($T_c$) using $A=k_BT_c/4a$ where $k_B$ is the Boltzman constant. When the blocks $\mathbf{i}$ and $\mathbf{j}$ are made of different materials, the estimation of $J_{\mathbf{ij}}$ is very challenging. Therefore, the IEC is usually assumed to be some kind of average of bulk exchanges or it is treated as a free parameter.

Alternatively, the micromagnetic parameters can be evaluated from first principles electronic structure calculations within a general multiscale scheme for hysteresis modeling.\cite{Belashchenko,Antropov,Kazantseva,Atxitia,Aas} An advantage of this approach is that, in addition to the bulk parameters, it also allows for calculations of the variations of these parameters in the proximity of an interface. In particular, the IEC can be explicitly evaluated.\cite{Aas} Such calculations, however, are often very expensive. In addition, the accuracy of first principles methods may be currently not sufficient for certain classes of materials including important for permanent magnetism rare-earth intermetalics .

\section{Exchange spring magnets: Introductory overview}

The basic physics of exchange spring magnets can be explained using 1D micromagnetic models for hard-soft layered structures.\cite{Kneller,Skomski,Goto,Kronmuller2,Kronmuller3,Solzi,Solzi2,Guo} The key length scale is the HP domain wall width,\cite{Kneller} $\delta_h=\pi\sqrt{A_h/K_h}$ where $A_h$ and $K_h$ denote the exchange stiffness constant and the anisotropy density constant of the HP, respectively. Let's consider a superlattice structure with an in-plane HP anisotropy axis and a strong IEC.  Assuming a thick HP layer, we can distinguish three different magnetization reversal regimes based on the soft layer thickness, $L_s$.\cite{Kronmuller2,Solzi} (1) For $L_s < \delta_h$ both layers are rigidly coupled producing a nearly rectangular hysteresis loop with a large coercive field ($H_c$) given approximately by\cite{Skomski} 

\begin{equation}
\mu_0H_c=2\frac{L_hK_h+L_sK_s}{M_hL_h+M_sL_s}
\end{equation}

where $M_h$ ($M_s$) and $K_h$ ($K_s$) are the spontaneous magnetization and the anisotropy density constant of the hard (soft) phase, respectively. As argued in Ref. \onlinecite{Skomski} the corresponding maximum energy product can be as high as 1 MJ/m$^3$, thus, making nanocomposites in this regime ideal for permanent magnet applications. (2) For $L_s \gtrsim \delta_h$ we are in the exchange spring regime. Here, the soft layer remains roughly parallel to the hard layer for reverse fields smaller than the nucleation field, which assuming a rigid hard layer can be written as\cite{Kronmuller4}

\begin{equation}
\mu_0H_{n}=\frac{2K_s}{M_s} + \frac{2\pi^2 A_s}{M_sL_s^2}
\end{equation}

\begin{figure}[t!]
\begin{center}
\includegraphics[width=1.0\hsize]{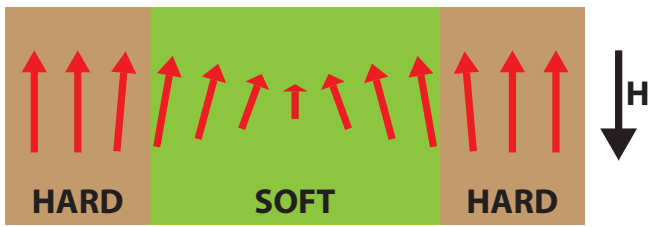}
\end{center}
\caption{A schematic picture describing a magnetic structure of hard-soft superlattice in the exchange spring regime when the reversed field exceeds the nucleation field. Red arrows represent the direction of layered resolved magnetization. The soft layer magnetic moments rotate away from the HP magnetization direction in order to decrease the Zeeman energy. The rotation angle increases with the increasing distance from the interface.}
\label{ExchangeSpring}
\end{figure}

The first term is the anisotropy field of the SP, while the second accounts for the exchange field from the hard layer.\cite{Goto} Once the reverse field exceeds $H_{n}$, a domain wall-like magnetic structure develops inside the soft layer. Here, the magnetic moments continuously rotate toward the direction of the external field with the rotation angle increasing as the distance from the hard layers increases (see Fig. \ref{ExchangeSpring}). The corresponding demagnetization curve has a convex shape and is fully reversible so that when the external field is removed the magnetization of the soft layer rotates back into alignment with the HP in analogy with the behavior of a mechanical spring. Further increase of the reverse field results in an increase of the rotation angles inside the soft layer until the depinning field is reached at which the domain wall propagates into the hard layer and the magnetization switches. (3) For $L_s \gg \delta_h$ both layers are essentially decoupled. The soft layer switches irreversibly at a low reverse field $\sim\mu_0H^{K}_{s}=2K_s/M_s$ creating a domain wall at each interface. On the other hand, the hard layer remains roughly intact until the reverse field is strong enough to reverse its magnetization. This reversal field, however, is not equal to the hard layer anisotropy field $\mu_0H^{K}_{h}=2K_h/M_h$, but it is the depinning field at which the domain wall at the interface starts to propagate inside the hard layer. Assuming infinitely thick hard and soft layers, the depinning field was derived to be\cite{Kronmuller3}

\begin{equation}
\mu_0H_{dp}=2\frac{K_hA_h-K_sA_s}{M_hA_h+M_sA_s+2\sqrt{M_sM_hA_sA_h}}
\label{Depin}
\end{equation}

In the special case of $K_s=0$, $M_s=M_h$, and $A_s=A_h$ the above formula simplifies to\cite{Aharoni} $H_{dp}=H^{K}_{h}/4$ demonstrating a coercivity decrease of a hard magnet due to an interface with the soft material. Note that a nonzero $K_s$, $M_s>M_h$, and $A_s>A_h$ all lead to further reduction of the depinning field. The above expression for the depinning field was found to be a good approximation as long as the interfacial domain walls fit inside each layer which makes Eq. (\ref{Depin}) applicable even in the exchange spring regime.\cite{Richter2,Suess}

The above discussion is applicable for hard-soft layered magnets with an in-plane anisotropy axis. In the case of the out-of-plane anisotropy axis there is another complication due to magnetic moments having components normal to the interface. This is due to magnetization discontinuity between the hard and soft layers leading to the creation of magnetic charges at the interface. The resulting stray field can be approximately taken into account by introducing effective anisotropy constants\cite{Solzi,Chantler,Solzi3,Bo}

\begin{equation}
K^{eff}_{h,s} = K_{h,s} - \mu_0M_{h,s}^2/2
\end{equation}

where the second term is the stray field energy contribution. Note that for a soft layer with small $K_s$ the effective anisotropy constant may become negative resulting in a reversal starting already in the first quadrant of the hysteresis loop.

The role of the IEC between the hard and soft materials was also studied.\cite{Solzi3,Bo,Zhao,Deng} It was demonstrated that the reduction of the IEC resulted in the decrease of the soft layer thickness for which hysteretic behavior changes from a rigid magnet to the exchange-spring regime.\cite{Zhao,Deng} In the exchange-spring regime the nucleation field increases with IEC, but the dependence is rather weak unless the IEC is an order of magnitude lower than the bulk exchange.\cite{Bo,Zhao,Deng} On the other hand, the depinning field increases as the IEC decreases. Similarly as for the nucleation field, however, the dependence is strong only when IEC is significantly smaller from the bulk exchange.\cite{Solzi3,Bo,Zhao,Deng}

Experimental realization of hard-soft magnetic nanocomposites requires control of the SP dimension at the nanometer scale. The first bulk exchange spring magnet was reported by Coehoorn\emph{et al.},\cite{Coehoorn} in 1989. Here, a Nd$_2$Fe$_{14}$B/Fe$_3$B composite with nanoscale isotropic grains was obtained using the melt-spinning technique. Subsequently, a number of hard-soft nanostructured magnets were produced\cite{Ding,Manaf,Krause,Zeng,Sort,Rong,Rui,Rong2} using various methods including melt-spinning, mechanical alloying, and nanoparticle self-assembly. The bulk exchange spring magnets, however, tend to be isotropic. Consequently, while their measured energy products are improved over corresponding isotropic single-phase magnets, they are smaller than the values obtained for corresponding oriented single-phase magnets. Therefore, much of the experimental focus was on thin film systems where the crystallographic texture can be controlled more easily.\cite{Liu,Fullerton,Zhang,Liu2,Sawatzki,Neu,Cui} In particular, the energy product enhancement as compared to the corresponding oriented single-phase magnets have been reported for FePt(L1$_0$)/Fe-Pt(fcc),\cite{Liu2} SmCo$_5$/Fe,\cite{Sawatzki,Neu} and Nd$_2$Fe$_{14}$B/FeCo\cite{Cui} nanocomposite films. 

While the simple 1D micromagnetic models capture basic physics of exchange-spring systems, they cannot provide a quantitative description of real magnets. In particular, the predicted energy products\cite{Skomski} are significantly larger than the experimental values. More accurate theoretical modeling of hard-soft nanocomposites can be obtained using numerical micromagnetic simulations.\cite{Berkov,Fidler,Kronmuller} This method takes into account a realistic 3D microstructure of magnetic systems and explicitly treats the long-range dipolar interactions. A number of calculations have been performed for different exchange-spring magnet systems\cite{Fisher,Fisher2,Fisher3,Schrefl,Schrefl2,Schrefl3,Kuma,Fukunaga, Fukunaga2,Chui,Griffiths,Rong3, Fukunaga3,Belemuk,Belemuk2,Belemuk3,Belemuk4,Belemuk5, Russier,He,Chen,Belemuk6, Fukunaga4,Saiden} In particular, it was demonstrated that permanent magnet properties of hard-soft nanocomposites are reduced as compared to the prediction of  1D micromagnetic models due to several mechanisms including grain misorientation, grain shape irregularity, and dipolar interaction induced magnetic inhomogeneities.

3D micromagnetic simulations result in a stronger dependence of the hysteretic properties on the IEC as compared to the 1D micromagnetic models.\cite{Fukunaga} However, it was argued that such dependence is an artifact of the micromagnetic approach which underestimates the exchange energy associated with the large changes of magnetization at the interface.\cite{Sanchez,Sanchez2,Sanchez3} In fact, the multiscale model which explicitly treats the interfacial atomic structure yields results very similar to the 1D models.\cite{Sanchez,Sanchez2,Sanchez3,Sanchez4}

\section{Results and discussion}

We investigated hard-soft composites with different nanostructure patterns including layered and core-shell geometries. The spontaneous magnetization of the SP was set to $\mu_0M_s=$2 T which is close to the spontaneous magnetization value for typical soft magnets like Fe (2.15 T), Co (1.8 T), and FeCo (2.45 T). On the other hand, the spontaneous magnetization values differ significantly between different hard materials. In particular, we have Nd$_2$Fe$_{14}$B (1.61 T), Sm$_2$Fe$_{17}$N$_3$ (1.54 T),  FePt-L1$_0$ (1.43 T) Sm$_2$Co$_{17}$ (1.28 T), SmCo$_5$ (1.07 T), CoPt-L1$_0$ (1.0 T), MnBi (0.78 T), and BaFe$_{12}$O$_{19}$ (0.47 T). In order to cover this wide range of values and to understand how the HP spontaneous magnetization affects the hysteretic behavior, in our simulations we vary this parameter from 0.1 T to 2 T. For the SP the anisotropy density constant was set to zero while for the HP we typically use $K_h=$5 MJm$^{-3}$ which is close to the room temperature value for Nd$_2$Fe$_{14}$B (4.9 MJm$^{-3}$). However, in order to study the effect of this parameter on the properties, some calculations are done for $K_h=$2.5 MJm$^{-3}$ or $K_h=$10 MJm$^{-3}$. The exchange stiffness for both the hard and SP was assumed to have a typical value of 10 pJ/m. The size of the micromagnetic block was set to 1 nm.

\subsection{Layered structures}

\begin{figure}[t!]
\includegraphics[width=1.0\hsize]{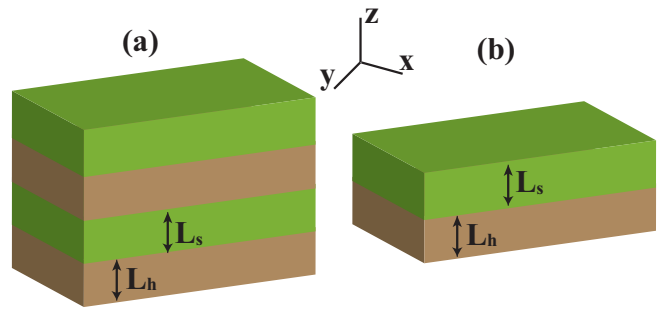}
\caption{Layered composite geometries. (a) Superlattice structure: 3D PBC are assumed. (b) Bilayer structure: 2D PBC are assumed in the $xy$ plane and the system is finite along the $z$ direction. In both case the sizes of the supercell along the $x$ and $y$ directions were set to 20 nm.}
\label{LayeredGeometries}
\end{figure}

Two different types of layered geometries are depicted in Fig. \ref{LayeredGeometries}. In both cases, PBC are assumed along the in-plane directions ($x$ and $y$ axes) and the size of the supercell along the $x$ and $y$ directions was set to 20 nm. For the superlattice structure (Fig. \ref{LayeredGeometries}a) PBC are also assumed along the out-of-plane direction so that the hard and soft layers periodically alternate along the $z$ axis (the supercell contains one hard and one soft layer). Alternatively, for the bilayer structure (Fig. \ref{LayeredGeometries}b) the system is finite in the out-of-plane direction and the hard-soft bilayer is surrounded by vacuum along the $z$ axis.

\begin{figure}[t!]
\includegraphics[width=1.0\hsize]{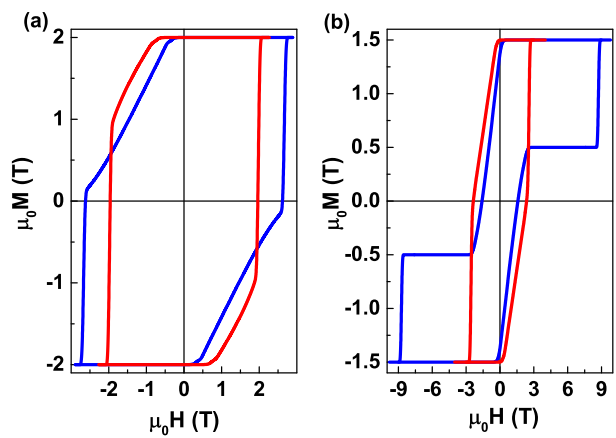}
\caption{Calculated hysteresis loops for the hard-soft superlattice structure with $L_h=L_s=$10 nm and $K_h=$5 MJ/m$^3$. The HP easy axis points along the out-of-plane direction. (a) $M_h=M_s=$2 T; (b) $M_h=M_s/2=$1 T. Red and blue lines correspond to $J_{\text{int}}=J_{\text{bulk}}$ and $J_{\text{int}}=0.2J_{\text{bulk}}$, respectively.}
\label{Hysteresis}
\end{figure}

Fig. \ref{Hysteresis} shows calculated hysteresis loops for the superlattice structure with $L_h=L_s=$10 nm and $K_h=$5 MJ/m$^3$. The easy axis of the HP was assumed to point along the out-of-plane direction. The $M_h=M_s=$2 T case (Fig. \ref{Hysteresis}a) represents a typical exchange-spring behavior. The magnetization starts deviating from its saturation value once the nucleation field is reached. This deviation is, however, reversible since the removal of the external field results in a return to the perfectly aligned magnetic state. As the reverse field keeps increasing, the magnetization continuously decreases until the depinning field is reached at which the magnetization irreversibly switches onto the field direction. For the strong IEC ($J_{\text{int}}=J_{\text{bulk}}$) the reversible region of the demagnetization curve has a convex shape but for the weak IEC ($J_{\text{int}}=0.2J_{\text{bulk}}$) the curve become more linear. In agreement with Refs. \onlinecite{Bo,Zhao,Deng} the reduction of the IEC results in a decrease of the nucleation field and an increase of the depinning field. 

For $M_h=M_s/2=$1 T (Fig. \ref{Hysteresis}b) we see that the reduction of the HP spontaneous magnetization leads to a decrease of the nucleation field. In fact, in the case of a weak IEC ($J_{\text{int}}=0.2J_{\text{bulk}}$) the nucleation field becomes negative with the reversal starting already in the first quadrant of the hysteresis loop. In addition, the reduction of the IEC leads to a much stronger increase of the depinning field.

\begin{figure}[t!]
\includegraphics[width=1.0\hsize]{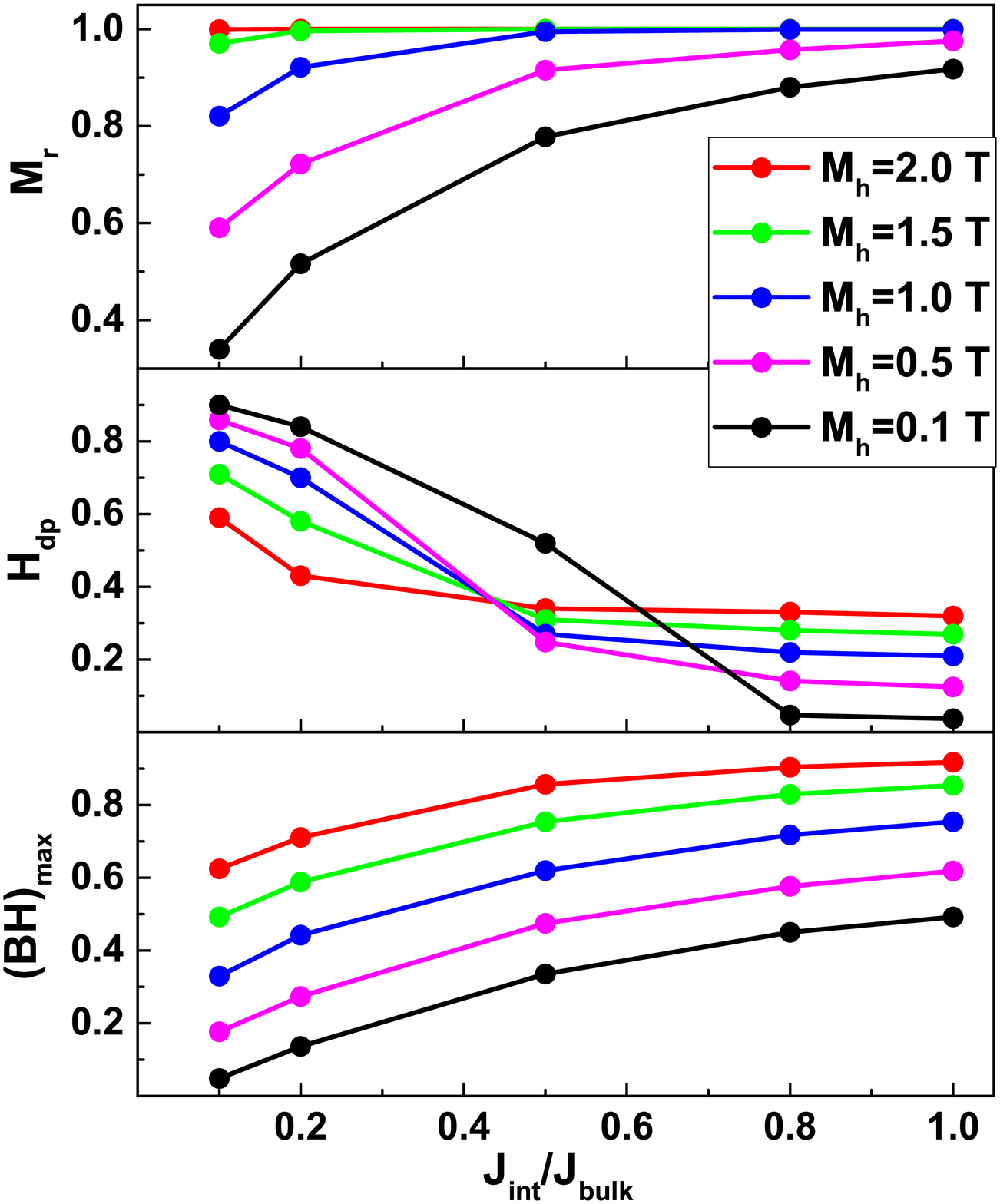}
\caption{Calculated hysteretic properties for the hard-soft superlattice structure as a function of the IEC ($J_{\text{int}}$) and the HP magnetization ($M_h$). We used $L_h=L_s=10$ nm, $M_s=$2 T, and $K_h=$5 MJ/m$^3$. The HP easy axis points along the out-of-plane direction. (top) Remanence in units of saturation magnetization ($M_{\text{sat}}=(L_hM_h+L_sM_s)/(L_h+L_s)$); (middle) Depinning field in units of the HP anisotropy field ($H_h=2K_h/(\mu_0M_h)$); (bottom) Maximum energy product in units of ($\mu_0M^2_{\text{sat}}/4$).}
\label{Properties}
\end{figure}

In order to more clearly illustrate how $M_h$ and $J_{\text{int}}$ influence the hysteretic properties we plotted the remanence, the depinning field, and the maximum energy product ($(BH)_{max}$) as a function of the IEC for different values of the HP spontaneous magnetization, see Fig. \ref{Properties}. We used $L_h=L_s=$10 nm and $K_h=$5 MJ/m$^3$. The remanence is in units of saturation magnetization ($M_{\text{sat}}=(L_hM_h+L_sM_s)/(L_h+L_s)$), the depinning field is in units of the HP anisotropy field ($H^{K}_h=2K_h/(\mu_0M_h)$), and $(BH)_{max}$ is in units of the ideal energy product ($\mu_0M^2_{\text{sat}}/4$). As seen for large values of $M_h$, the remanence is close to the saturation magnetization and shows a very weak dependence on the IEC. However, when $M_h$ becomes significantly lower  than the SP spontaneous magnetization, the remanence is reduced from the saturation magnetization value and shows a much stronger decrease as the IEC gets smaller. 

The depinning field increases as $M_h$ decreases mainly because the coupling of the HP magnetic moments to the external field becomes weaker (see Eq. \ref{Zeeman}). However, when this trivial effect is taken into account by expressing $H_{dp}$ in units of $H^{K}_h$, we see that for a strong IEC $H_{dp}/H^{K}_h$ is an increasing function $M_h$. The depinning field  increases under IEC reduction since the propagation of the magnetic reversal from the soft to HP is more difficult when the interaction between both phases is diminished. The dependence of $H_{dp}$ on the IEC becomes stronger as $M_h$ decreases. As a result, for weak IEC $H_{dp}/H^{K}_h$ decreases with $M_h$. 

The maximum energy product increases with $M_h$ partially because of the $M_{\text{sat}}$ increase and partially because of the enhancement of the remanence to saturation ratio. Reduction of the IEC diminishes $(BH)_{max}$ due to the decrease of the remanent magnetization and despite the increase of the depinning field. Similarly as for the remanence and $H_{dp}$, the dependence on the IEC becomes stronger as $M_h$ gets smaller.

\begin{figure}[t!]
\includegraphics[width=1.0\hsize]{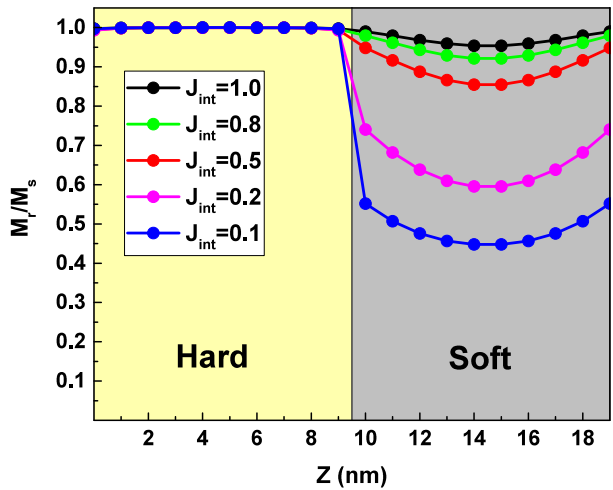}
\caption{Layer-resolved magnetization (component along the $z$ axis) in the remanent state for the superlattice structure as a function of the position along the superlattice axis ($z$) for different values of the IEC ($J_{int}$, in units of the bulk exchange). The magnetization is normalized to the value spontaneous magnetization for a given $z$ ($M_h$ or $M_s$). The HP easy axis points in the out-of-plane direction. Hard and soft layer thicknesses were both set to 10 nm. We used $M_h=M_s/4=$0.5 T and $K_h=$5 MJ/m$^3$.}
\label{LayerResM}
\end{figure}

In order to better understand the above results, the layer-resolved magnetization in the remanent state was plotted as a function of the position along the superlattice axis for different values of the IEC (see Fig. \ref{LayerResM}). Hard and soft layer thicknesses were both set to 10 nm. We assumed $M_h=M_s/4=$0.5 T and $K_h=$5 MJ/m$^3$. As seen, the magnetic moments inside the hard layer are aligned with the HP easy axis ($z$ axis). However, the soft layer magnetic moments are tilted away from the $z$ axis with the tilting angle increasing with the distance from the hard-soft interface. This magnetic structure is similar to the one in Fig. \ref{ExchangeSpring} except that the HP magnetization is perpendicular to the interface. We point out, however, that the external field is zero here and the magnetic inhomogeneity is instead driven  by the dipolar interaction. Indeed, in the perfectly aligned state there is a magnetization discontinuity at the hard-soft interface which leads to creation of interfacial magnetic charges and increase of the system magnetostatic energy. In order to remove this discontinuity, the SP magnetic moments tend to tilt away from the HP magnetization direction.\cite{Belemuk,Belemuk4}  Note that as the difference between $M_s$ and $M_h$ increases, the magnitude of the interfacial magnetic charges increases resulting in a stronger tilting. This explains the reduction of the remanence to the saturation ratio when $M_h/M_s$ decreases. More importantly, however, the tilting competes with the IEC which prefers the SP and HP magnetizations to be aligned. Thus, the interface magnetic charges are not completely removed and the degree of tilting depends on the strength of the IEC. This mechanism explains the enhancement of the dependence of the remanence on the IEC when $M_h/M_s$ is reduced. 

The behavior of the depinning field can also be understood. For a strong IEC the presence of the  magnetization discontinuity at the interface diminishes $H_{dp}$ since the domain wall propagation into the HP reduces the discontinuity. Therefore, the depinning field increases as $M_h/M_s$ approaches unity. On the other hand, for a weak IEC the magnetic charges are removed by the SP tilting. This tilting configuration acts as an additional pinning center suppressing the domain wall motion into the HP. Thus, $H_{dp}$ increases as $M_h/M_s$ decreases. It follows that the dependence of $H_{dp}$ on IEC becomes stronger as the difference between $M_s$ and $M_h$ increases.

Note that the dipolar interaction induced SP tilting exists only for the HP easy axis pointing in the out-of-plane direction. If the easy axis is in the in-plane direction, the magnetic charges are not created at the interface even for a small $M_h/M_s$ ratio. In this case, the remanence to saturation ratio (not shown) is always equal to unity showing no dependence on $M_h$ and IEC.

\begin{figure}[t!]
\includegraphics[width=1.0\hsize]{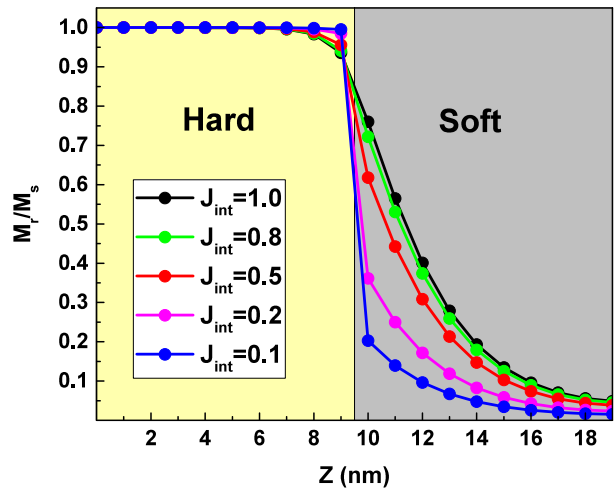}
\caption{Layer-resolved magnetization (component along the $z$ axis) in the remanent state for the bilayer structure as a function of the position along the bilayer axis ($z$) for different values of the IEC ($J_{int}$, in units of the bulk exchange). The magnetization is normalized to the value spontaneous magnetization for a given $z$ ($M_h$ or $M_s$). The HP easy axis points in the out-of-plane direction. Hard and soft layer thicknesses were both set to 10 nm. We used $M_h=M_s/4=$0.5 T and $K_h=$5 MJ/m$^3$.}
\label{LayerResMBilayer}
\end{figure}

\begin{figure}[t!]
\includegraphics[width=1.0\hsize]{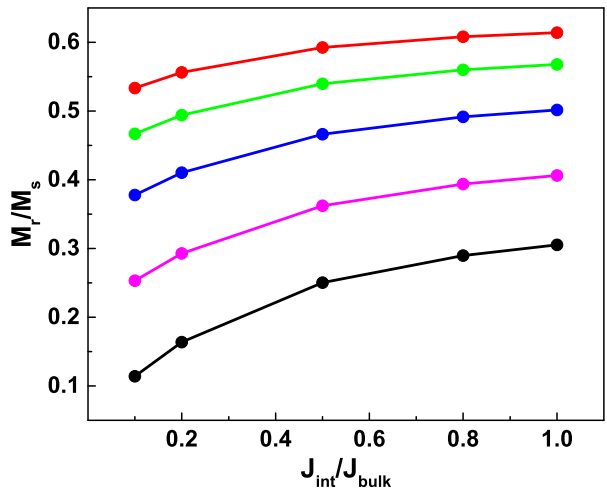}
\caption{Remanent magnetization for the hard-soft bilayer structure as a function of the IEC.  Red, green, blue, magenta, and black circles correspond to $M_h=$2 T, $M_h=$1.5 T, $M_h=$1 T, $M_h=$0.5 T, and $M_h=$0.1 T, respectively. We used $L_h=L_s=10$ nm, $M_s=$2 T, and $K_h=$5 MJ/m$^3$. The HP easy axis points along the out-of-plane direction. The remanence is in units of saturation magnetization ($M_{\text{sat}}=(L_hM_h+L_sM_s)/(L_h+L_s)$)).}
\label{RemanenceBilayer}
\end{figure}

Let's consider now the bilayer structure (Fig. \ref{LayeredGeometries}b). Here, in addition to the hard-soft interface, both materials also have a free surface at which magnetic charges can be created, as long as the the HP easy axis points in the out-of-plane direction. Fig. \ref{LayerResMBilayer} shows the layer-resolved magnetization in the remanent state as a function of the position along the bilayer axis for different values of the IEC. Similar to the superlattice case we assume $L_h=L_s=$ 10 nm, $M_h=M_s/4=$0.5 T, and $K_h=$5 MJ/m$^3$. As seen, the strong anisotropy of the hard layer aligns all of its magnetic moments along the easy axis even at the cost of creating magnetic charges at the hard layer surface. Alternatively, the soft layer magnetic moments tilt away from the $z$ axis in order to avoid magnetization discontinuity at the interface and at the SP surface. The tilt angle increases as we go away from the interface and reaches almost 90 degrees at the surface. The remanent magnetization as a function of the IEC for different values of $M_h$ is shown in Fig. \ref{RemanenceBilayer}. Even when the HP and SP spontaneous magnetizations are equal, the soft layer magnetic moments still tilt in order to remove magnetic charges at the soft surface resulting in a significant reduction of the remanence. In particular, for a strong IEC the remanence to saturation ratio is only 0.6 and decreases further when the IEC is reduced. For smaller $M_h$ the additional tendency to remove magnetic charges at the hard-soft interface enhances the tilting resulting in even stronger decrease of the remanence to saturation ratio.

\begin{figure}[t!]
\includegraphics[width=1.0\hsize]{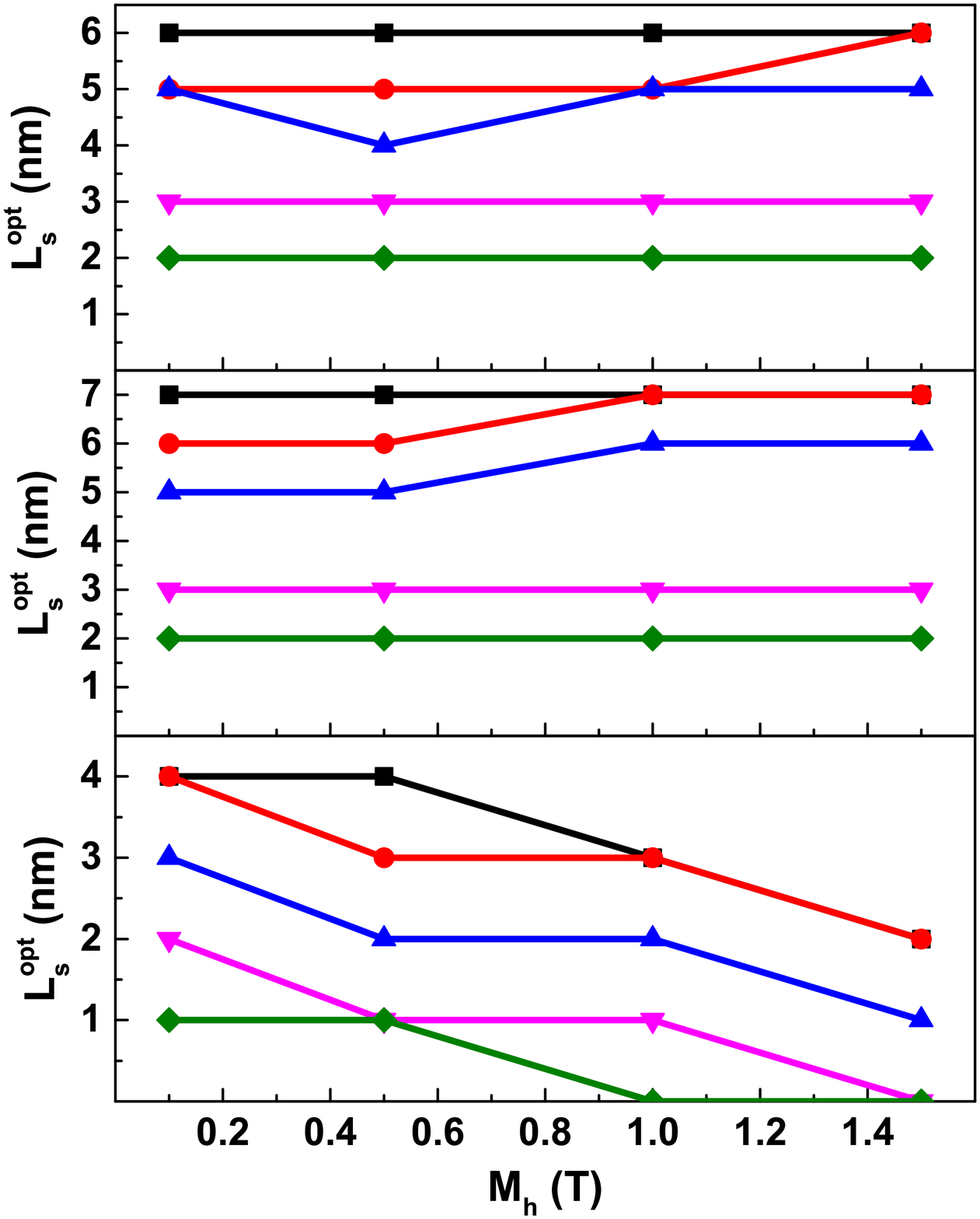}
\caption{The optimal thickness of the soft layer ($L_s^{\text{opt}}$) for the superlattice structure as a function of the HP magnetization ($M_h$) for different values of the IEC. The HP easy axis points along the out-of-plane direction. Black, red, blue, magenta, and green circles correspond to $J_{\text{int}}/J_{\text{buk}}=$1.0, 0.8, 0.5, 0.2, and 0.1, respectively. The hard layer thickness was set to 10 nm. (top) $K_h=$2.5 MJ/m$^3$; (middle) $K_h=$5 MJ/m$^3$; (bottom) $K_h=$10 MJ/m$^3$.}
\label{OptimalLs}
\end{figure}

An important figure of merit for hard-soft nanocomposites is the optimal size of the SP for which the energy product is maximized. This quantity can be roughly estimated as twice that of the HP domain wall width,\cite{Kneller} $\delta_h=\pi\sqrt{A_h/K_h}$. However, its exact value may depend on many factors including nanocomposite geometry, the HP spontaneous magnetization, the HP anisotropy density constant, as well as the IEC. In Fig. \ref{OptimalLs} we show the calculated optimal soft layer thickness ($L_s^{\text{opt}}$) for the superlattice structure with an out-of-plane easy axis as a function of $M_h$ for different strengths of IEC and different values of the HP anisotropy constant. As long as the hard layer is not too thin to allow for tunneling of reversal modes,\cite{SkomskiBook} $L_s^{\text{opt}}$ should not depend on the hard layer thickness. Here we set $L_h=$10 nm. By comparing three panels in Fig. \ref{OptimalLs} we observe that $L_s^{\text{opt}}$ is a nonmonotonic function of $K_h$. In particular, let's consider the case of strong IEC ($J_{\text{int}}=J_{\text{bulk}}$). For $K_h=$10 MJ/m$^3$ the optimal soft layer thickness is equal to 6 nm which is close to $2\delta_h\approx$6.3 nm. When $K_h$ decreases to 5 MJ/m$^3$, $L_s^{\text{opt}}$ increases to 7 nm which can be explained by the increase of the HP domain wall width ($2\delta_h\approx$8.9 nm). Further decrease of $K_h$ to 2.5 MJ/m$^3$ results in $2\delta_h\approx$12.6 nm. However, the calculated $L_s^{\text{opt}}$ dramatically decreases to a value between 2 and 4 nm depending on $M_h$. The reason for this discrepancy is the fact that the $L_s^{\text{opt}}\approx2\delta_h$ estimate is valid only for large $K_h$. Indeed, while the estimate is very accurate for $K_h=$10 MJ/m$^3$, for $K_h=$5 MJ/m$^3$ it is off by almost 2 nm and for $K_h=$2.5 MJ/m$^3$ it is completely wrong. Physically, the reason for the failure of this estimate is that for small $K_h$ the domain wall is not pinned at the interface but propagates inside the hard layer. Therefore, one should be careful when using the $L_s^{\text{opt}}\approx2\delta_h$ approximation for the hard-soft nanocomposites with a moderate HP anisotropy (e.g., MnBi).

The optimal soft layer thickness decreases when the IEC  is diminished since the reduction of the coupling  between hard and soft layers makes the exchange hardening mechanism less effective. In particular, for $K_h=$10 MJ/m$^3$ the decrease of $J_{\text{int}}$ from $J_{\text{bulk}}$ to 0.2$J_{\text{bulk}}$ results in the reduction of $L_s^{\text{opt}}$ by a factor of two. 

\begin{figure}[t!]
\includegraphics[width=1.0\hsize]{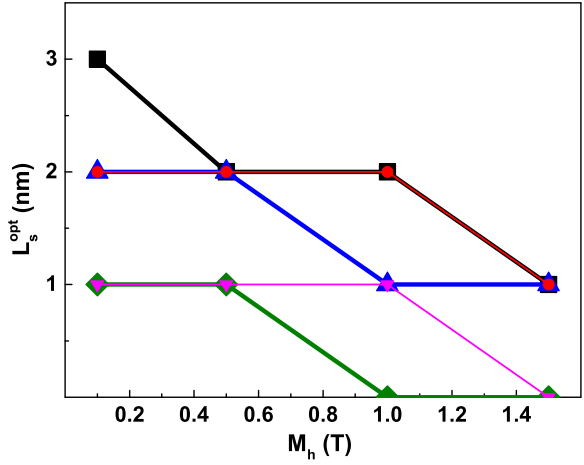}
\caption{The optimal thickness of the soft layer ($L_s^{\text{opt}}$) for the bilayer structure as a function of the HP magnetization ($M_h$) for different values of the IEC. The HP easy axis points along the out-of-plane direction. Black, red, blue, magenta, and green shapes correspond to $J_{\text{int}}/J_{\text{buk}}=$1.0, 0.8, 0.5, 0.2, and 0.1, respectively.  We used $L_h=$10 nm and $K_h=$5 MJ/m$^3$.}
\label{OptimalLsBilayer}
\end{figure}

The dependence of $L_s^{\text{opt}}$ on the HP magnetization is complicated. On one hand, a decrease of $M_h$ leads to a larger gain in the magnetization and the energy product when the SP is added. Therefore, the optimal soft layer thickness should increase. On the other hand, when the $M_h/M_s$ ratio is reduced, the magnetization and the energy product decreases due to the aforementioned tilting of the soft layer magnetic moments. This results in a decrease of $L_s^{\text{opt}}$. Which of these two mechanisms prevails depends on other parameters. In particular, for a strong HP anisotropy ($K_h\geq$5 MJ/m$^3$) the latter mechanism is only slightly more important and consequently $L_s^{\text{opt}}$ shows a small increase as $M_h$ gets larger. For $K_h\geq$2.5 MJ/m$^3$ the former mechanism prevails, resulting in $L_s^{\text{opt}}$ being a decreasing function of $M_h$.

Fig. \ref{OptimalLsBilayer} shows the optimal soft layer thickness for the bilayer structure with an out-of-plane easy axis as a function of $M_h$ for different values of IEC. The anisotropy density constant was set to $K_h=$5 MJ/m$^3$. In general, $L_s^{\text{opt}}$ is smaller by more than a factor of two as compared to the superlattice structure. This is partially caused by the fact that in the bilayer case only one hard-soft interface is present and, therefore, the exchange hardening mechanism is reduced by half. In addition, the tilting of the soft layer magnetic moments is more pronounced for the bilayer geometry in order to remove magnetic charges at the soft surface. As in the superlattice case, a decrease of the IEC results in a smaller $L_s^{\text{opt}}$. However, the optimal soft layer thickness for the bilayer consistently increases as $M_h$ decreases. Indeed, even when $M_h=M_s$, the soft layer magnetic moments tilt in order to avoid magnetization discontinuity at the soft layer surface. While the decrease of the HP magnetization increases the tilting (see Fig. \ref{LayerResMBilayer}), this effect is less important than the increase of the magnetization gain under the soft layer addition when the $M_h/M_s$ ration is reduced. Therefore, $L_s^{\text{opt}}$ is a decreasing function of $M_h$.

\begin{figure}[t!]
\begin{center}
\includegraphics[width=1.0\hsize]{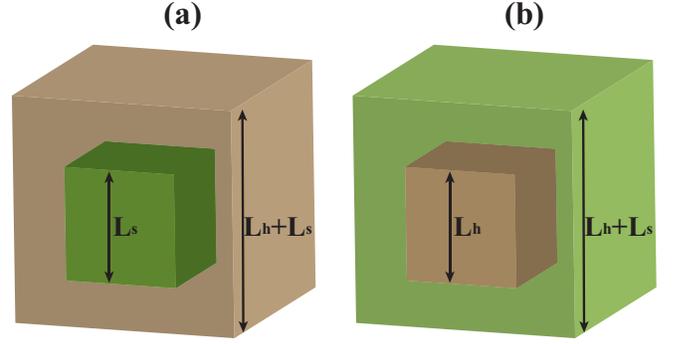}
\end{center}
\caption{Core-shell composite geometries. (a) SP cubic inclusion inside the HP matrix. (b) HP cubic inclusion inside the SP matrix. 3D PBC are assumed. The inclusions are uniformly distributed within the matrix phase. In the case of the soft (hard) inclusion the separation between the inclusions is denoted by $L_h$ ($L_s$).}
\label{CoreShellGeometries}
\end{figure}

\subsection{Core-shell structures}

Two types of core-shell geometry are presented in Fig. \ref{CoreShellGeometries}. The first structure is a SP cube of side length $L_s$ surrounded by a HP shell, see Fig. \ref{CoreShellGeometries}a. 3D PBC are assumed and therefore this geometry corresponds to uniformly distributed SP cubic inclusions inside the HP matrix. The separation between the inclusions is denoted by $L_h$. The second structure is a HP cube of side length $L_h$ surrounded by a SP shell, see Fig. \ref{CoreShellGeometries}b. 3D PBC are again assumed so that the system corresponds to uniformly distributed HP cubic inclusions inside the SP matrix. The separation between the inclusions is denoted by $L_s$.

\begin{figure}[t!]
\includegraphics[width=1.0\hsize]{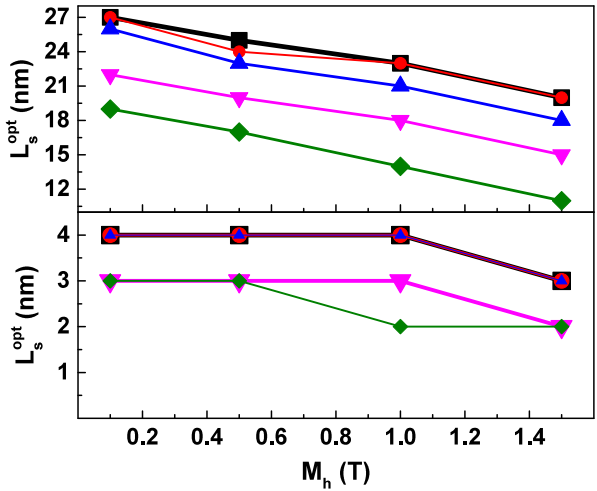}
\caption{The optimal SP size which maximizes $(BH)_{\text{max}}$ for the core-shell geometries. (top) The optimal side length of the SP cube ($L_s^{\text{opt}}$) for the geometry from Fig. \ref{CoreShellGeometries}a s a function of $M_h$ for different values of the IEC. The separation between inclusions was set to $L_h=$5 nm. (bottom) The optimal separation between the HP inclusions ($L_s^{\text{opt}}$) for the geometry from Fig. \ref{CoreShellGeometries}b s a function of $M_h$ for different values of the IEC. The side length of the HP cube was set to $L_h=$6 nm. For both structures we used $K_h=$5 MJ/m$^3$. Black, red, blue, magenta, and green circles correspond to $J_{\text{int}}/J_{\text{buk}}=$1.0, 0.8, 0.5, 0.2, and 0.1, respectively.}
\label{OptimalLsCoreShell}
\end{figure}

Figure \ref{OptimalLsCoreShell} shows the optimal size of the SP component which maximizes $(BH)_{\text{max}}$ for the core-shell geometries. For both structures $K_h=$5 MJ/m$^3$ was used. For the HP matrix the separation between the inclusions was set to $L_h=$5 nm while for the SP matrix the inclusion size was assumed to be $L_s=$6 nm. In the case of SP inclusions the optimal SP size is much larger than in the case of layered geometries. The reason for this is that the soft layers have reduced size only along one direction and are extended in two other directions. On the other hand, inclusions have reduced size along all three directions which allows for a very effective exchange hardening. On the other hand, in the case of HP inclusions the optimal size of the SP is reduced as compared to the layered structures since the soft component is extended in all directions.

Similarly as for layered structures, for both core-shell geometries $L_s^{\text{opt}}$ decreases when the IEC is diminished. On the other hand, a reduction of $M_h$ leads to an increase of $L_s^{\text{opt}}$ even though an opposite trend was observed for superlattices with the out-of-plane easy HP axis and the same anisotropy constant ($K_h=$5 MJ/m$^3$). This indicates that the tendency of the SP magnetic moments to tilt in order to remove magnetic charges at the hard-soft interface is less weaker for the core-shell structure. Indeed, for both geometries we found that the remanence to saturation ratio is almost independent on $M_h$ and roughly equal to one. The reason for this is that while there are always two interface perpendicular to the HP easy axis, the soft regions adjacent to these interfaces are connected to the SP regions adjacent to the interfaces parallel to the HP easy axis. Therefore, the tilting is unfavorable by the exchange interaction. However, as demonstrated in Refs. \onlinecite{Belemuk3}, at finite temperatures the tilting can become significant due to thermal fluctuations.

The lack of a significant tilting for the core-shell geometry seems to disagree with results of Ref. \onlinecite{Belemuk7} where it was demonstrated that for a SmFeN inclusion inside the FeCo shell the perfectly aligned state is unstable with respect to the tilting for $L_{FeCo}\geq$ 6.6 nm. We point out, however, that the PBC in Ref. \onlinecite{Belemuk7} correspond to a spherical sample surrounded by vacuum. Therefore, the aligned magnetic state is disfavored by a nonzero demagnetization field. In this work, on the other hand, closed circuit PBC are assumed and the demagnetization field is zero.

\section{Conclusions}

In summary, we introduced a micromagnetic approach for calculations of magnetic hysteresis. The method allows for PBC along arbitrary number of directions. The long-range dipolar interactions are evaluated using the Ewald summation technique which was described for 3D, 2D, and 1D PBC. For a given system the Ewald construction needs to be done only once and, therefore, its contribution to the computational time is negligible. The method was applied to investigate demagnetization behavior of hard-soft magnetic nanocomposites. Different geometries were considered including layered and core-shell structures. 

We studied the dependence of the hysteretic properties on the strength IEC. For layered geometries with an out-of-plane HP easy axis we found the demagnetization process is very sensitive to the strength of the IEC, as long as the the spontaneous magnetization for the HP is significantly lower than for the SP. We demonstrated that this is caused by an interplay of the IEC and the dipolar interactions. The latter promote the tilting of the SP magnetization away from the HP easy axis direction in order to avoid creation of the magnetic charges at the interface.\cite{Belemuk,Belemuk4} This tendency, however, competes with the IEC which prefers alignment of HP and SP magnetizations. As a result, the strength of the IEC is crucial for the permanent magnet properties of the composites.

The SP tilting is absent for layered geometries when the HP easy axis is in the in-plane direction. In the case of core-shell structures the magnetic charges can be created at interfaces perpendicular to the HP easy axis (two per inclusion). However, the core-shell geometry has a specific topology for which SP regions adjacent to these interfaces are connected to the SP regions adjacent to the interfaces parallel to the HP easy axis. Therefore, the tilting is suppresses by the exchange interaction inside the SP. As a result, for layered geometries with an in-plane HP easy axis and the core-shell structures the hysteretic properties have a weak dependence on the IEC (as long as the IEC is of the same order as the bulk exchange coupling).

Further, we studied the effect of micromagnetic parameters and geometry on the optimal SP ($L_s^{\text{opt}}$) size that leads to the largest $(BH)_{\text{max}}$ of the composite. We found that the well known approximation\cite{Kneller} $L_s^{\text{opt}}\approx2\delta_h$ is valid only if the HP anisotropy is large enough. In particular, we showed that the approximation is reasonably accurate for $K_h\geq$5 MJ/m$^3$. In this regime $L_s^{\text{opt}}$ increases as $K_h$ decreases. However, for smaller anisotropy values ($K_h=$2.5 MJ/m$^3$) the approximation breaks down and reduction of $K_h$ results in a decrease of $L_s^{\text{opt}}$. 

For the core-shell geometry with SP inclusions the optimal SP size is much larger than in the case of layered geometries due to the fact that the inclusions have a reduced size along all directions. However, in the case of HP inclusion inside the SP matrix $L_s^{\text{opt}}$ is reduced as compared to the layered structures since the SP component is extended in all directions.

In general, the optimal SP thickness decreases when the IEC is reduced due to diminished role of the exchange hardening mechanism. On the other hand, reduction of $M_h$ (with respect to $M_s$) typically increases $L_s^{\text{opt}}$ since addition of the SP leads then to a larger magnetization gain. However, when the SP tilting induced by the magnetization disontinuity at the hard-soft interface is important (superlattice structures with out-of-plane easy axis and strong anisotropy), the decrease of $M_h$ enhances the tilting and reduced the magnetization resulting in a decrease of $L_s^{\text{opt}}$.

\section*{Acknowledgments}
This work was supported by the project: 'Solid State Processing of Fully Dense Anisotropic Nanocomposite Magnets', ARPA-E Control Number 0670-4987, and by the Critical Materials Institute, an Energy Innovation Hub funded by the U.S. Department of Energy (DOE), and by the Office of Basic Energy Science, Division of Materials Science and Engineering. The research was performed at Ames Laboratory, which is operated for the U.S. DOE by Iowa State University under contract \# DE-AC02-07CH11358.

\appendix*

\section{}

In this appendix we describe how to evaluate the effective dipolar tensor for 1D, 2D, and 3D PBC using the Ewald technique. The elements of the effective dipolar tensor can be written as

\begin{equation}
\bar{D}_{\mathbf{i}\mathbf{j}}^{\alpha\beta}\equiv\bar{D}_{\alpha\beta}(\mathbf{r})=\frac{\partial^2}{\partial r_\alpha\partial r_\beta}\Phi(\mathbf{r})
\label{deriv}
\end{equation}

where we defined $\mathbf{r}=(x,y,z)\equiv\mathbf{r}_{\mathbf{ij}}$ and

\begin{equation}
\Phi(\mathbf{r})=\sum_{\mathbf{n}}{}^{'}\frac{1}{|\mathbf{r}+\mathbf{n}|}
\end{equation}

Here, $\mathbf{n}=(n_x,n_y,n_z)$ where $n_\alpha$ is an integer enumerating the periodic images of the supercell if PBC are assumed along the direction $\alpha$ and zero otherwise. The prime at the summation excludes the $\mathbf{n}=\mathbf{0}$ term when $\mathbf{r}=\mathbf{0}$. In the Ewald method we write $\Phi(\mathbf{r})$ in the following way

\begin{equation}
\Phi(\mathbf{r})=\Phi^{(\text{r})}(\mathbf{r})+\Phi^{(\text{k})}(\mathbf{r})+\Phi^{(\text{si})}(\mathbf{r})
\label{phi}
\end{equation}

The first two terms are given by

\begin{eqnarray}
\Phi^{(\text{r})}(\mathbf{r})=\sum_{\mathbf{n}}{}^{'}\frac{\text{erfc}\left(\eta|\mathbf{r}+\mathbf{n}|\right)}{|\mathbf{r}+\mathbf{n}|} \\
\label{phir}
\Phi^{(\text{k})}(\mathbf{r})=\sum_{\mathbf{n}}\frac{\text{erf}\left(\eta|\mathbf{r}+\mathbf{n}|\right)}{|\mathbf{r}+\mathbf{n}|}
\label{phik}
\end{eqnarray}

where $\text{erf}(x)$ and $\text{erfc}(x)$ are error and complementary error functions, respectively. The parameter $\eta$ is arbitrary and can be chosen in such a way to make the calculations the most effective (see below). The last term in Eq. (\ref{phi}) is $\Phi^{(\text{si})}(\mathbf{r})=-\frac{\text{erf}(\eta r)}{r}$ for $\mathbf{r}=\mathbf{0}$ and zero otherwise. This self-interaction accounts for the lack of the prime at the sum in the expression for $\Phi^{(\text{k})}$.

The differentiation in Eq. (\ref{deriv}) leads to three corresponding contributions to the effective dipolar tensor

\begin{equation}
\bar{D}_{\alpha\beta}(\mathbf{r})=\bar{D}^{\text{(r)}}_{\alpha\beta}(\mathbf{r})+\bar{D}^{\text{(k)}}_{\alpha\beta}(\mathbf{r})+\bar{D}^{\text{(si)}}_{\alpha\beta}(\mathbf{r})
\end{equation}

Evaluation of the self-interaction term $\bar{D}^{\text{(si)}}$ results in

\begin{equation}
\bar{D}_{\alpha\beta}^{\text{(si)}}(\mathbf{r})=\delta_{\mathbf{r0}}\delta_{\alpha\beta}\frac{4}{3}\frac{\eta^3}{\sqrt{\pi}}
\label{dsi}
\end{equation}

where $\delta_{ij}$ is the Kronecker delta function. The real space contribution is given by

\begin{align}
\nonumber
\bar{D}^{\text{(r)}}_{\alpha\beta}(\mathbf{r})=\sum_{\mathbf{n}}{}^{'}\big[\left(r_{\alpha}+n_{\alpha}\right)\left(r_{\beta}+n_{\beta}\right)C\left(|\mathbf{r}+\mathbf{n}|\right) \\
 - \delta_{\alpha\beta}B\left(|\mathbf{r}+\mathbf{n}|\right)\big]
\label{dr}
\end{align}

where

\begin{eqnarray}
B\left(r\right)=\big[\frac{2\eta r}{\sqrt{\pi}}e^{-\eta^2r^2}+ \text{erfc}(\eta r) \big]/r^3 \\
\nonumber
C\left(r\right)=\big[\frac{2\eta r}{\sqrt{\pi}}\left(3+2\eta^2r^2\right)e^{-\eta^2r^2} \\ 
+ 3\text{erfc}(\eta r) \big]/r^5
\end{eqnarray}

The complimentary error function in $\Phi^{(\text{r})}$  makes the sum in Eq. (\ref{dr}) to be quickly convergent so that $\bar{D}^{\text{(r)}}$ can be easily evaluated numerically. 

Thanks to the error function in Eq. (\ref{phik}) $\Phi^{(\text{k})}(\mathbf{r})$ is a smooth functions of $\mathbf{r}$ so that it can be efficiently expanded into a Fourier series. In order to do this we need to specify the dimensionality of the PBC. Below we consider all three cases.

\textbf{3D PBC:} Since the system is periodic in all three directions we can write

\begin{equation}
\Phi_{\text{3D}}^{(\text{k})}(\mathbf{r})=\frac{1}{NV}\sum_{\mathbf{G}}c_{\mathbf{G}}e^{i\mathbf{G}\cdot\mathbf{r}}
\end{equation}

where $N$ is the number of periodic images, $V=L_xL_yL_z$ is the system volume, and $\mathbf{G}=2\pi\left(\frac{m_x}{L_x},\frac{m_y}{L_y},\frac{m_z}{L_z}\right)$ with $m_x,m_y,m_z$ being integers. The Fourier coefficients are equal to

\begin{equation}
c_{\mathbf{G}}=\sum_{\mathbf{n}}\int_V\frac{\text{erf}\left(\eta|\mathbf{r}+\mathbf{n}|\right)}{|\mathbf{r}+\mathbf{n}|}e^{-i\mathbf{G}\cdot\mathbf{r}}d\mathbf{r}
\end{equation}

The $\mathbf{G}=\mathbf{0}$ coefficient is set to zero since it corresponds to the macroscopic demagnetizing field which vanishes under the assumption of closed-circuit environment. For $\mathbf{G}\neq\mathbf{0}$ we obtain 

\begin{equation}
c_{\mathbf{G}}=\frac{4\pi N}{G^2}\exp\left[-\left(\frac{G}{2\eta}\right)^2\right]
\end{equation}

resulting in

\begin{equation}
\Phi_{\text{3D}}^{(\text{k})}(\mathbf{r})=\frac{4\pi}{V}\sum_{\mathbf{G}\neq\mathbf{0}}\frac{e^{i\mathbf{G}\cdot\mathbf{r}}}{G^2}\exp\left[-\left(\frac{G}{2\eta}\right)^2\right]
\end{equation}

and

\begin{align}
\nonumber
\bar{D}^{\text{(k)3D}}_{\alpha\beta}(\mathbf{r})=-\frac{4\pi}{V}\sum_{\mathbf{G}\neq\mathbf{0}}\frac{G_{\alpha}G_{\beta}}{G^2}\exp\left[-\left(\frac{G}{2\eta}\right)^2\right] \\
\times\cos(\mathbf{G}\cdot\mathbf{r})
\end{align}

The exponential factor makes the above sum quickly convergent and easy to evaluate numerically. Typically, $\eta=5/(L_xL_yL_z)^{1/3}$ is chosen.

\textbf{2D PBC:} We assume that PBC are applied along the $x$ and $y$ axes and there is no periodicity along the $z$ axis. Expanding $\Phi_{\text{2D}}^{(\text{k})}$ into a 2D Fourier series we obtain

\begin{equation}
\Phi_{\text{2D}}^{(\text{k})}(\mathbf{r})=\frac{1}{NS}\sum_{\mathbf{G}}c_{\mathbf{G}}e^{i\mathbf{G}\cdot\mathbf{r}_{xy}}
\end{equation}

where $N$ is the number of periodic images, $S=L_xL_y$ is the area of the system perpendicular to the $z$ axis, $\mathbf{r}_{xy}=(r_x,r_y)$ is the projection of the $\mathbf{r}$ vector on the $xy$ plane, and $\mathbf{G}=2\pi\left(\frac{m_x}{L_x},\frac{m_y}{L_y}\right)$ is the 2D reciprocal lattice vector with $m_x,m_y$ being integers. The Fourier coefficients are equal to

\begin{equation}
c_{\mathbf{G}}=\sum_{\mathbf{n}}\int_S\frac{\text{erf}\left(\eta|\mathbf{r}+\mathbf{n}|\right)}{|\mathbf{r}+\mathbf{n}|}e^{-i\mathbf{G}\cdot\mathbf{r}_{xy}}d\mathbf{r}_{xy}
\end{equation}

In the case of 2D PBC the $\mathbf{G}=\mathbf{0}$ coefficient is nonzero and equal to

\begin{equation}
c_{\mathbf{G}=\mathbf{0}}=-2N\left[\frac{\sqrt{\pi}}{\eta}\exp\left(-\eta^2r_z^2\right) + \pi z\text{erf}(\eta r_z)  \right]
\end{equation}

For $\mathbf{G}\neq\mathbf{0}$ we get

\begin{equation}
c_{\mathbf{G}}=\frac{\pi N}{G}\left[D(r_z)+D(-r_z)\right]
\end{equation}

where we defined

\begin{equation}
D(r_z)=e^{Gr_z}\text{erfc}\left(\frac{G}{2\eta}+\eta r_z\right)
\end{equation}

We then obtain

\begin{widetext}
\begin{equation}
\Phi_{\text{2D}}^{(\text{k})}(\mathbf{r})=-\frac{2}{S}\left[\frac{\sqrt{\pi}}{\eta}\exp\left(-\eta^2r_z^2\right) + \pi z\text{erf}(\eta r_z)  \right] +\frac{\pi}{S}\sum_{\mathbf{G}\neq\mathbf{0}}\frac{e^{i\mathbf{G}\cdot\mathbf{r}_{xy}}}{G} \left[D(r_z)+D(-r_z)\right]
\end{equation}

and

\begin{align}
\nonumber
\bar{D}^{\text{(k)2D}}_{\alpha\beta}(\mathbf{r})=-\frac{\pi}{S}\delta_{\alpha z}\delta_{\beta z}\sum_{\mathbf{G}}\left(E(r_z)-G\left[D(r_z)+D(-r_z)\right]\right)\cos(\mathbf{G}\cdot\mathbf{r}_{xy}) -\frac{\pi}{S}\sum_{\mathbf{G}\neq\mathbf{0}}[\delta_{\alpha z}\delta_{\beta,xy}G_{\beta}+\delta_{\beta z}\delta_{\alpha,xy}G_{\alpha}] \\
\times\left[D(r_z)-D(-r_z)\right]\sin(\mathbf{G}\cdot\mathbf{r}_{xy})-\frac{\pi}{S}\delta_{\alpha,xy}\delta_{\beta,xy}\sum_{\mathbf{G}\neq\mathbf{0}}\frac{G_{\alpha}G_{\beta}}{G}\left[D(r_z)+D(-r_z)\right]\cos(\mathbf{G}\cdot\mathbf{r}_{xy})
\end{align}
\end{widetext}

where $\delta_{\alpha,\beta\gamma}=(\delta_{\alpha\beta}+\delta_{\alpha\gamma})/2$ and 

\begin{equation}
E(r_z)=\frac{4\eta}{\sqrt{\pi}}\exp\left(-\frac{G^2}{4\eta^2}-\eta^2r_z^2\right)
\end{equation}

The sums over $\mathbf{G}$ are quickly convergent and can be readily evaluated numerically. Typically, $\eta=5/\sqrt{L_xL_y}$ is chosen.

\textbf{1D PBC:} We assume that PBC are applied along the $x$ axis and there is no periodicity along the $y$ and $z$ axes. Expanding $\Phi_{\text{1D}}^{(\text{k})}(\mathbf{r})$ into 1D Fourier series we obtain

\begin{equation}
\Phi_{\text{1D}}^{(\text{k})}(\mathbf{r})=\frac{1}{NL_x}\sum_{G}c_{G}e^{iGr_x}
\end{equation}

where $N$ is the number of periodic images, and $G=2\pi\frac{m_x}{L_x}$ with $m_x$ being an integer. The Fourier coefficients are equal to

\begin{equation}
c_{G}=\sum_{\mathbf{n}}\int_0^{L_x}\frac{\text{erf}\left(\eta|\mathbf{r}+\mathbf{n}|\right)}{|\mathbf{r}+\mathbf{n}|}e^{-iGr_x}dr_x
\label{cG1D}
\end{equation}

The $G=0$ coefficient is equal to

\begin{align}
c_{G=0}=-N \Big[ \gamma + \Gamma\left( 0,\eta^2r^2_{yz}\right) +\log\left(\eta^2r^2_{yz} \right) \Big]
\end{align}

where $\mathbf{r}_{yz}=(r_y+r_z)$ is the projection of the $\mathbf{r}$ vector on the $yz$ plane, $\gamma$ is the Euler constant, and $\Gamma(s,t)$ is the incomplete gamma function. For $G\neq 0$ the integral (\ref{cG1D}) cannot be obtained in a closed form. Following the approach from Ref. \onlinecite{Porto} we therefore express the solution as the power series in $r^2_{yz}$ 

\begin{equation}
c_{G}=N\sum_{k=0}^{\infty}\frac{(-1)^k}{4^kk!}G^{2k}r^{2k}_{yz}\Gamma\left(-k,\frac{G^2}{4\eta^2}\right)
\end{equation}

The above is only applicable for $r^2_{yz}\neq0$. In the special case $r^2_{yz}=0$ we have instead

\begin{equation}
c_{G}=N\Gamma\left(0,\frac{G^2}{4\eta^2}\right)
\end{equation}

We thus obtain

\begin{widetext}
\begin{equation}
\Phi_{\text{1D}}^{(\text{k})}(\mathbf{r})=-\frac{1}{L_x} \left[ \gamma + \Gamma\left( 0,\eta^2r^2_{yz}\right) + \log\left(\eta^2r^2_{yz} \right) \right] +\frac{1}{L_x}\sum_{G\neq0}e^{iGr_x}\begin{cases}
\sum\limits_{k=0}^{\infty}\frac{(-1)^k}{4^kk!}G^{2k}r^{2k}_{yz}\Gamma\left(-k,\frac{G^2}                    {4\eta^2}\right)&\text{if } r^2_{yz}\neq0 \\
\Gamma\left(0,\frac{G^2}{4\eta^2}\right)&\text{if } r^2_{yz}=0
\end{cases}
\end{equation}

The effective dipolar matrix for $r_{yz}^2\neq0$ is given by

\begin{align}
\nonumber
\bar{D}^{\text{(k)1D}}_{\alpha\beta}(\mathbf{r})=\frac{2\delta_{\alpha,yz}\delta_{\beta,yz}}{L_xr_{yz}^2}\left[ \delta_{\alpha\beta}\left(e^{-\eta^2r_{yz}^2}-1\right) -2r_{\alpha}r_{\beta}\left(\eta^2e^{-\eta^2r_{yz}^2} + \frac{e^{\eta^2r_{yz}^2}-1}{r_{yz}^2}\right) \right] -\frac{\delta_{\alpha x}\delta_{\beta x}}{L_x}\sum_{G}e^{iGr_x}\sum_{k=0}^{\infty}\\
\nonumber
\times \frac{(-1)^k}{4^kk!}G^{2(k+1)}r_{yz}^{2k}\Gamma\left(-k,\frac{G^2}{4\eta^2}\right) + \frac{4}{L_x}\delta_{\alpha,yz}\delta_{\beta,yz}r_{\alpha}r_{\beta}\sum_{G\neq0}e^{iGr_x}\sum_{k=0}^{\infty}\frac{k(k-1)(-1)^k}{4^kk!}G^{2k}r_{yz}^{2(k-2)}\Gamma\left(-k,\frac{G^2}{4\eta^2}\right) \\
+\frac{2i}{L_x}\left(\delta_{\alpha x}\delta_{\beta,yz}r_{\beta}+\delta_{\alpha,yz}\delta_{\beta x}r_\alpha\right)\sum_{G}e^{iGr_x}\sum_{k=0}^{\infty}\frac{k(-1)^k}{4^kk!}G^{2k+1}r_{yz}^{2(k-1)}\Gamma\left(-k,\frac{G^2}{4\eta^2}\right)
\end{align}

while for $r_{yz}^2=0$ we have

\begin{equation}
\bar{D}^{\text{(k)1D}}_{\alpha\beta}(\mathbf{r})=-\frac{1}{L_x}\left[\delta_{\alpha x}\delta_{\beta x}\sum_{G}G^2e^{iGr_x}\Gamma\left(0,\frac{G^2}{4\eta^2}\right) + \frac{1}{2}\delta_{\alpha,yz}\delta_{\beta,yz}\sum_{G}G^2e^{iGr_x}\Gamma\left(-1,\frac{G^2}{4\eta^2}\right) + 2\eta^2\delta_{\alpha\beta}\delta_{\alpha,yz}\right]
\end{equation}

The numerical evaluations of $\bar{D}^{\text{(k)1D}}$ can be straightforwardly done. Typically, $\eta=5/L_x$ is used.

\end{widetext}

\end{document}